\documentclass[12pt]{iopart}

\usepackage{iopams}
\usepackage{graphicx}
\begin{document}

\title{Dynamical properties of the Rabi model}

\author{Binglu Hu$^{1}$, Huili Zhou$^{1}$, Shujie Chen$^{1}$, Gao Xianlong$^{1}$, and Kelin Wang$^{2}$}

\address{$^{1}$ Department of Physics, Zhejiang Normal University, Jinhua
321004, China}
\address{$^{2}$ Department of Modern Physics, University of Science and Technology of China,
Hefei 230026, China}

\begin{abstract}
We study the dynamical properties of the quantum Rabi model within a systematic expansion method. 
Based on the observation that the parity symmetry of the Rabi model is kept during the evolution of the states,
we decompose the initial state and the time-dependent one into a part of a positive and a negative parity expanded by
the superposition of the coherent states. The evolutions for the corresponding positive and the negative parity are obtained, 
where the expansion coefficients in the dynamical equations are known from the recurrence relation derived.
\end{abstract}

\section{Introduction}
The interplay of light and matter is vital for the understanding on the magnetic field on an atom possessing a nuclear spin. Semi-classical and quantum-mechanical methods are usually used to investigate phenomena involving light and its interactions with matter. Semi-classically, the radiation field is treated as a classical field and the interactions between the matter and the radiation field are considered as certain interaction terms in the Hamiltonian.
Quantum mechanically, the light is treated as quantized photons~\cite{books}. The simplest model which takes the full quantum-mechanical treatment
of the interaction of the radiation field with the matter is the quantum Rabi model~\cite{Rabi36}, where the matter is a simplest two-level atom. 
Nowadays, this model is realized and applied in a variety of quantum systems, from quantum optics, condensed-matter physics to ultracold atomic systems. 
A few examples includes optical-cavity quantum electrodynamics (QED)~\cite{cavity QED,haroche}, trapped ions~\cite{Dots}, quantum dots, and circuit QED~\cite{circuit QED,wallraff1}. Due to the development in ultracold atomic systems~\cite{Bloch}, the precise simulation and control of the quantum Rabi model becomes possible, ranging from the weak, strong, ultrastrong, to deep strong coupling regimes.

Recently the dynamical nature of the quantum-Rabi-like model has been addressed~\cite{Schweber, Larson, Nataf, Schiro, Wolf, Liberato, Henriet}. 
For the quantum Rabi model in the rotating-wave approximation (RWA) (by neglecting the counter rotating terms corresponding to virtual light-matter excitations), i.e., the celebrated Jaynes-Cummings model~\cite{JC}, the dynamical description is given which predicts collapses and revivals of the population inversion. 
Dynamical correlation functions of the quantum Rabi model, through time-dependent correlation functions, are studied to classify the the ultrastrong and deep strong coupling of the quantum Rabi model~\cite{Wolf}. 
A stochastic Schr\"{o}dinger equation approach was developed to access the strong-coupling limit of the Rabi model and study the effects of dissipation and ac drive in an exact manner~\cite{Henriet}.
The advances in ultracold atomic systems make the experimental study on the dynamical properties possible~\cite{Bloch}.
A quantum simulation of the quantum Rabi model with cold atoms was proposed by loading the atoms onto a periodic lattice~\cite{Felicetti}. The effective two-level quantum system is simulated by different Bloch bands in the first Brillouin zone, and the bosonic radiation field is represented by the motion of the atomic cloud in a superimposed harmonic optical-trap potential. Hereby the time evolution of a quantum system prepared in an initial state, can be simulated by a quantum quench. In this way, the dynamical properties of the quantum Rabi model are reached in the extreme parameter regime, for example, in the following mentioned ultrastrong and deep strong coupling regime, which is the key interest in the present study of the quantum Rabi model.

In this paper, we will propose a systematic method to investigate the dynamical properties of the quantum Rabi model beyond the adiabatic approximation and RWA. The rest of the paper is organized as follows: in Sec. II, we introduce the model. In Sec. III, we explicitly derive the dynamic equations for the positive and negative parity and thus the recurrence equations. The conclusion is given in the last section. 

\section{The model}
The quantum Rabi model, considers a two-level atom (or spin 1/2) of transition frequency $\Delta$ coupled to a single-mode cavity field with mode frequency $\omega$ (described by a quantized harmonic oscillator $\omega a^\dagger a $), which is a minimal model for the interaction of matter and light,
\begin{equation}
\label{HamilRabi}
H = \case\Delta2  \sigma_z + \omega a^\dagger a + \lambda \sigma_x (a+a^\dagger).
\end{equation}
Here $\sigma_{x,y,z}$ are the Pauli matrices, and $a^\dagger$ and $a$ the bosonic creation and annihilation operators of the cavity field. The coupling strength between light and atom is $\lambda$. The three parameters $\Delta$, $\omega$ and $\lambda$ determines all the static and distinct dynamics. We set $\hbar=1$ throughout the whole paper.

The ratio of the coupling strength $\lambda$ and the mode frequency $\omega$, $g=\lambda/\omega$, determines the different properties of the model and restricts the corresponding theoretical methods. The typical quantum optical values give $g\sim 10^{-6}$, and the present circuit QED gives $g \sim 10^{-2}$. This is the weak-coupling regime, where the RWA applies and the continuous $U(1)$-symmetry results in the exactly solvable Jaynes-Cummings model~\cite{JC,RWA}. 
Nowadays, the superconducting circuits has reached the so-called ultrastrong (UTS) coupling regime with $g\sim 0.1$~\cite{Ciuti,Gunter}. Recently, a deep strong coupling regime with $g \ge 1$ is considered, where the dynamics can be explained as photon number wavepacket propagating along a defined parity chains of the Hilbert space~\cite{Casanova}. In the latter two cases, the RWA fails and more elegant methods demand.

Among them, analytical solutions of the quantum Rabi model beyond the RWA have been recently explored~\cite{Braak,Chen,Zhong,Maciejewski,cont,jonasBerry}. It has recently been shown that the Rabi model undergoes a second-order QPT in the joint limit of $\Delta/\omega\rightarrow \infty$ and $\lambda\rightarrow \infty$ where the control parameter $ 2\lambda/\sqrt{\omega\Delta}$ is kept constant~\cite{Hwang15}. The results show that the quantum Rabi model has a discrete $Z_2$ (parity) symmetry~\cite{Braak,Hwang10}, that is, $[P,H]=0$, where the parity operator $P=\sigma_z e^{i\pi a^{\dagger} a}$ satisfying $P\vert p\rangle=p\vert p\rangle$ and $p=\pm 1$. The Hilbert space splits into two infinite-dimensional invariant space, the positive ($+$) and negative ($-$) parity parts. 
The following dynamical evolution of the Rabi model will be built from these two Hilbert space.

\section{Dynamical evolution of the Rabi model}

In the following, we will show that how the system dynamics evolves inside the Hilbert space split in two unconnected subspaces or parity chains.

The time-dependent problem can be solved either through a time-dependent Schr\"{o}dinger equation $i\partial_t \vert t\rangle=H\vert t\rangle$ by evolving a state vector $\vert t\rangle=e^{-iH (t-t_0)}\vert t_0\rangle$
or a Heisenberg equation $\partial_t \rho(t)=-i [H, \rho(t)]$ by evolving a state density matrix $\rho(t)=e^{-iH (t-t_0)}\rho(t_0)e^{iH (t-t_0)}$. 
For both cases, the numerical simulation of the state evolution has to be done through the functional integral via Trotterization of the time interval,
\[
e^{iH (t-t_0)}={\rm lim}_{N\rightarrow \infty} (1+i\delta_t H)^N, ~~\delta_t =\frac{t-t_0}{N}.
\]
To simplify the numerical calculation of the Rabi model, an adiabatic approximation can be taken by substituting a full Hamiltonian $H$ into an adiabatic one $H_{\rm adia}$~\cite{Larson} and $\vert t\rangle=e^{-iH_{\rm adia} (t-t_0)}\vert t_0\rangle$. 
The validity of the adiabatic approximation is judged from a full wavepacket simulation by the split operator procedure~\cite{Fleit}, which relies on the separation of the evolution operator. 
For the open system in the UTS regime, the second-order time-convolutionless projection operator method is used to correctly describe the dynamics of the system~\cite{Forndiaz} and $\vert t\rangle=e^{-iH_{\rm UTS} (t-t_0)}\vert t_0\rangle$.
So in many cases, the evolutions of the wavepacket are given using perturbation theory or computed numerically.

Here we consider the dynamical properties for the system preparing in some initial state $\vert t=0\rangle$ evolving according to the full quantum mechanical Hamiltonian. A generic quench protocol can realize this procedure~\cite{Felicetti,Pedernales}. We should mention that the initial state is not necessarily an eigenstate of Eq. (\ref{HamilRabi}). At $t >0$, the system unitarily evolves with the Hamiltonian $H$,
\begin{eqnarray}
\vert t \rangle &= e^{-iH t}\vert t=0 \rangle \nonumber\\
&= \vert  t=0\rangle -iHt\vert  t=0\rangle +\frac{1}{2!} (-iH)^{2}t^{2} \vert  t=0\rangle +\ldots \nonumber\\
&~~~+\frac{1}{j!} (-iH)^{j}t^{j} \vert  t=0\rangle +\ldots~.
\label{eq:evolve}
\end{eqnarray}

Due to the parity symmetry of the Rabi model, the system keeps the same parities during the evolution of the states. Therefore the state vector at any time can be divided into a positive and a negative parity,
\begin{equation}
\vert  t \rangle =\vert +,t\rangle +\vert -,t\rangle .
\label{eq:t}
\end{equation}
It is also true for any initial state, $\vert t=0\rangle$, 
\begin{equation}
\vert  t=0 \rangle =\vert +,0\rangle +\vert -,0\rangle .
\label{eq:t0}
\end{equation}

By substituting Eqs. (\ref{eq:t}) and (\ref{eq:t0}) into Eq. (\ref{eq:evolve}), we obtain alternatively the time-dependent states for the even and odd parities, respectively, 
\begin{equation}
\vert+,t\rangle =\vert+,0\rangle -iH t \vert+,0\rangle-\frac{1}{2}H^{2}t^{2}\vert+,0\rangle +\ldots,
\label{eq:upc}
\end{equation}
\begin{equation}
\vert-,t\rangle =\vert-,0\rangle -iHt\vert-,0\rangle-\frac{1}{2}H^{2}t^{2}\vert-,0\rangle +\ldots.
\label{eq:dnc}
\end{equation}
The states $\vert+,t\rangle $ and $\vert-,t\rangle $ can be represented by the two component spinors which are further assumed as the superposition of the coherent state,
\begin{equation}
\vert +,t\rangle =
\left(
\begin{array}{c}
\int\int \psi_{1}(\zeta,\eta; t)(\vert\zeta+i\eta\rangle -\vert -\zeta-i\eta\rangle ) \mathrm{d}\zeta\mathrm{d}\eta\\
\int\int \psi_{2}(\zeta,\eta; t)(\vert\zeta+i\eta\rangle +\vert -\zeta-i\eta\rangle ) \mathrm{d}\zeta\mathrm{d}\eta
\end{array}
\right),
\label{eq:coherent1}
\end{equation}
\begin{equation}
\vert -,t\rangle =
\left(
\begin{array}{c}
\int\int \varphi_{1}(\zeta,\eta; t)(\vert\zeta+i\eta\rangle +\vert -\zeta-i\eta\rangle ) \mathrm{d}\zeta\mathrm{d}\eta\\
\int\int \varphi_{2}(\zeta,\eta; t)(\vert\zeta+i\eta\rangle -\vert -\zeta-i\eta\rangle ) \mathrm{d}\zeta\mathrm{d}\eta
\end{array}
\right),
\label{eq:coherent2}
\end{equation}
where $\psi_{1,2}$ and $\varphi_{1,2}$ are the corresponding expansion coefficients and $\vert \zeta+i\eta\rangle$ the coherent state defined as,
\begin{equation}
\vert \zeta+i\eta\rangle =\exp[(\zeta+i\eta)a^\dagger ]\vert 0 \rangle .
\end{equation}
As a result, to solve Eqs. (\ref{eq:upc}) and (\ref{eq:dnc}) is equivalent to deriving the recurrence formula for the time-dependent coefficients.
For this purpose, we need to make use of the operation of the bosonic operators $a, a^\dagger, a^\dagger a$ contained in the Hamiltonian $H$ on the coherent state $\vert \zeta+i\eta\rangle $. Thus, the relation between the Fock state $\vert n\rangle$ and the coherent state $\vert \zeta+i\eta\rangle$ is important (see the derivation in $\ref{Sec:Fockcoh}$),
\begin{eqnarray}
 \vert n\rangle 
&=\int\int e^{-\zeta^{2}-\eta^{2}}\sum_{l=0}^{n}\frac{\sqrt{n!}}{l!(n-l)!}(\zeta)^{n-l}(-i\eta)^{l}\vert\zeta+i\eta\rangle \frac{\mathrm{d}\zeta\mathrm{d}\eta}{\pi}.
\label{FockCoh}
\end{eqnarray} 

In order to calculate Eq. (\ref{eq:upc}), firstly we need to know what $H \vert+,t=0\rangle$, $H^2 \vert+,t=0\rangle$, and $H^j \vert+,t=0\rangle$ $(j\ge 3)$ are.
$\vert +,t=0\rangle $ is defined by Eq. ($\ref{eq:coherent1}$) at $t=0$. 
We set the result of $H^j$ ($j=1, 2, 3...$) on $\vert +,t=0\rangle $ is 
\begin{eqnarray}
H^j \vert +,0\rangle =
\left(
\begin{array}{c}
\int  \int A _{j}(\zeta ,\eta )[\vert  \zeta +i\eta \rangle -\vert  -\zeta -i\eta \rangle ] \frac{d\zeta d\eta }{\pi }\\
\int  \int B _{j}(\zeta ,\eta) [\vert  \zeta +i\eta\rangle +\vert  -\zeta -i\eta\rangle   ] \frac{d\zeta d\eta }{\pi } 
\end{array}
\right).
\label{eq:hofj}
\end{eqnarray}
By making use of the operation of the bosonic operators $a, a^\dagger, a^\dagger a$ on the coherent state derived in $\ref{Sec:oper}$, we firstly obtain the corresponding coefficients $A _{1}(\zeta ,\eta )$ and $B _{1}(\zeta ,\eta )$ in $H \vert+,t=0\rangle$,
\begin{eqnarray}
A _{1}(\zeta ,\eta )
&=\pi \frac{\Delta }{2}\psi _{1}(\zeta ,\eta; 0) +\lambda \pi \psi _{2}(\zeta ,\eta; 0)(\zeta +i\eta  )\nonumber\\
&+\sum^\infty_{m=0}\sum_{l=0}^{2m+1}\frac{2m+1}{l!(2m+1-l)!} \e^{-\zeta ^{2}-\eta ^{2}}\zeta ^{2m+1-l}(-i\eta )^{l}\int  \int  d\zeta _{1}d\eta _{1} \nonumber\\
&\times \left[  \lambda \psi _{2}(\zeta _{1},\eta _{1}; 0)(\zeta _{1}+i\eta _{1})^{2m}
+ \omega \psi _{1}(\zeta _{1},\eta _{1}; 0)(\zeta _{1}+i\eta _{1})^{2m+1} \right],
\label{eq:A1}
\end{eqnarray}
and
\begin{eqnarray}
B _{1}(\zeta ,\eta )
&=-\pi \frac{\Delta }{2}\psi _{2}(\zeta,\eta; 0) +\lambda \pi \psi _{1}(\zeta ,\eta ;0)(\zeta +i\eta  )\nonumber\\
&+\sum^\infty_{m=0}\sum_{l=0}^{2m+2}\frac{2m+2}{l!(2m+2-l)!}\e^{-\zeta ^{2}-\eta ^{2}}\zeta ^{2m+2-l}(-i\eta )^{l} \int  \int  d\zeta _{1}d\eta _{1} \nonumber\\
&\times \left[  \lambda \psi _{1}(\zeta _{1},\eta _{1};0)(\zeta _{1}+i\eta _{1})^{2m+1}
+ \omega \psi _{2}(\zeta _{1},\eta _{1};0)(\zeta _{1}+i\eta _{1})^{2m+2} \right].
\label{eq:B1}
\end{eqnarray}
Similarly, we can get the corresponding coefficients $A _{2}(\zeta ,\eta )$ and $B _{2}(\zeta ,\eta )$ in $H^2 \vert+,t=0\rangle$.
Comparing to Eqs. (\ref{eq:A1}) and (\ref{eq:B1}), what we need to do is to substitute $\psi _{1}(\zeta ,\eta ;t=0) \rightarrow A _{1}(\zeta ,\eta )$ and $\psi _{2}(\zeta ,\eta ;t=0)\rightarrow B _{1}(\zeta ,\eta )$. By repeating the above procedure, the recurrence formula are given for all the coefficients $A_{j}(\zeta ,\eta ) $ and $B _{j}(\zeta ,\eta )$ ($j=0, 1, 2, ...$),
\begin{eqnarray}
A _{j+1}(\zeta ,\eta )
&=\pi \frac{\Delta }{2}A _{j}(\zeta ,\eta ) +\lambda \pi B _{j}(\zeta ,\eta )(\zeta +i\eta  )\nonumber\\
&+\sum^\infty_{m=0}\sum_{l=0}^{2m+1}\frac{2m+1}{l!(2m+1-l)!} \e^{-\zeta ^{2}-\eta ^{2}}\zeta ^{2m+1-l}(-i\eta )^{l}\int  \int  d\zeta _{1}d\eta _{1} \nonumber\\
&\times \left[  \lambda B _{j}(\zeta_1 ,\eta_1 )(\zeta _{1}+i\eta _{1})^{2m}
+ \omega A _{j}(\zeta_1 ,\eta_1 ) (\zeta _{1}+i\eta _{1})^{2m+1} \right],
\label{eq:An}
\end{eqnarray}
and
\begin{eqnarray}
B _{j+1}(\zeta ,\eta )
&=-\pi \frac{\Delta }{2}B _{j}(\zeta ,\eta ) +\lambda \pi A _{j}(\zeta ,\eta )(\zeta +i\eta  )\nonumber\\
&+\sum^\infty_{m=0}\sum_{l=0}^{2m+2}\frac{2m+2}{l!(2m+2-l)!} \e^{-\zeta ^{2}-\eta ^{2}}\zeta ^{2m+2-l}(-i\eta )^{l}\int  \int  d\zeta _{1}d\eta _{1} \nonumber\\
&\times \left[  \lambda A _{j}(\zeta_1 ,\eta_1 )(\zeta _{1}+i\eta _{1})^{2m+1}
+ \omega B_{j}(\zeta_1 ,\eta_1 )(\zeta _{1}+i\eta _{1})^{2m+2} \right].
\label{eq:Bn}
\end{eqnarray}
Here we let $A_0(\zeta ,\eta)\equiv \psi _{1}(\zeta ,\eta ;t=0)$ and $B_0(\zeta ,\eta)\equiv \psi _{2}(\zeta ,\eta ;t=0)$.

In summary, given the initial states and thus the coefficients $\psi _{1}(\zeta ,\eta ;t=0)$ and $\psi _{2}(\zeta ,\eta ;t=0)$, we can obtain all the other coefficients $A_{n+1}$ and $B _{n+1}$
according to the recurrence relations $(\ref{eq:An})$ and $(\ref{eq:Bn})$, and thus, Eq. (\ref{eq:upc}) is solved in a systematic way,
\begin{eqnarray}
\vert  +,t\rangle &=
\left(
\begin{array}{c}
\int\int \psi _{1}(\zeta ,\eta ;0)[\vert  \zeta +i\eta \rangle -\vert  -\zeta -i\eta \rangle ]d\zeta d\eta \nonumber\\
\int \int\psi _{2}(\zeta ,\eta ;0)[\vert  \zeta +i\eta \rangle +\vert  -\zeta -i\eta \rangle ]d\zeta d\eta
\end{array}
\right)\nonumber\\
&-it
\left(
\begin{array}{c}
\int\int A _{1}(\zeta ,\eta)[\vert  \zeta +i\eta \rangle -\vert  -\zeta -i\eta \rangle ]d\zeta d\eta\\
\int \int B _{1}(\zeta ,\eta)[\vert  \zeta +i\eta \rangle +\vert  -\zeta -i\eta \rangle ]d\zeta d\eta
\end{array}
\right)
+\ldots\nonumber\\
&+\frac{(-i)^{j}}{j!}t^{j}
\left(
\begin{array}{c}
\int\int A _{j}(\zeta ,\eta)[\vert  \zeta +i\eta \rangle -\vert  -\zeta -i\eta \rangle ]d\zeta d\eta\\
\int \int B_{j}(\zeta ,\eta)[\vert  \zeta +i\eta \rangle +\vert  -\zeta -i\eta \rangle ]d\zeta d\eta
\end{array}
\right)
+\ldots~.
\label{eq:PP}
\end{eqnarray}

The same efforts can be done for Eq. (\ref{eq:dnc}) with,
\begin{eqnarray}
\vert -,t\rangle &=
\left(
\begin{array}{c}
\int \int\varphi  _{1}(\zeta ,\eta ;0)[\vert  \zeta +i\eta \rangle +\vert  -\zeta -i\eta \rangle ]d\zeta d\eta\nonumber\\
\int \int \varphi _{2}(\zeta ,\eta ;0)[\vert  \zeta +i\eta \rangle -\vert  -\zeta -i\eta \rangle ]d\zeta d\eta
\end{array}
\right)\nonumber\\
&-it
\left(
\begin{array}{c}
\int  \int C _{1}(\zeta ,\eta)[\vert  \zeta +i\eta \rangle +\vert  -\zeta -i\eta \rangle ]d\zeta d\eta \\
\int  \int D _{1}(\zeta ,\eta)[\vert  \zeta +i\eta \rangle -\vert  -\zeta -i\eta \rangle ]d\zeta d\eta
\end{array}
\right)
+\ldots\nonumber\\
&+\frac{(-i)^{j}}{j!}t^{j}
\left(
\begin{array}{c}
\int  \int C _{j}(\zeta ,\eta)[\vert  \zeta +i\eta \rangle +\vert  -\zeta -i\eta \rangle ]d\zeta d\eta\\
\int  \int  D  _{j}(\zeta ,\eta)[\vert  \zeta +i\eta \rangle -\vert  -\zeta -i\eta \rangle ]d\zeta d\eta
\end{array}
\right)
+\ldots~.
\label{eq:NP}
\end{eqnarray}
The coefficients are determined by the following recurrence relations for $C_j$ and $D_j$ with $j=0, 1, 2, ...$, 
\begin{eqnarray}
C _{j+1}(\zeta ,\eta )&=\pi \frac{\Delta }{2}C_{j}(\zeta ,\eta) +\lambda \pi D_{j}(\zeta ,\eta) (\zeta +i\eta  )\nonumber\\
&+ \sum^\infty_{m=0}\sum_{l=0}^{2m+2}\frac{2m+2}{l!(2m+2-l)!}\e^{-\zeta ^{2}-\eta ^{2}}\zeta ^{2m+2-l}(-i\eta )^{l}\int\int d\zeta _{1}d\eta _{1} \nonumber\\
&\times \left[ \lambda D_{j}(\zeta_1 ,\eta_1)(\zeta _{1}+i\eta _{1})^{2m+1}
+\omega C_{j}(\zeta_1 ,\eta_1)(\zeta _{1}+i\eta _{1})^{2m+2}\right ],
\label{eq:Cn}
\end{eqnarray}
and
\begin{eqnarray}
D _{j+1}(\zeta ,\eta )&=-\pi \frac{\Delta }{2}D_{j}(\zeta ,\eta) +\lambda \pi C_{j}(\zeta ,\eta)(\zeta +i\eta  )\nonumber\\
&+ \sum^\infty_{m=0}\sum_{l=0}^{2m+1}\frac{2m+1}{l!(2m+1-l)!} \e^{-\zeta ^{2}-\eta ^{2}}\zeta ^{2m+1-l}(-i\eta )^{l} \int\int d\zeta _{1}d\eta _{1} \nonumber\\
&\times \left[ \lambda C_{j}(\zeta _{1},\eta _{1})(\zeta _{1}+i\eta _{1})^{2m}
+\omega D_{j} (\zeta _{1},\eta _{1})(\zeta _{1}+i\eta _{1})^{2m+1}\right ].
\label{eq:Dn}
\end{eqnarray}
Here we let $C_0(\zeta ,\eta)\equiv \varphi _{1}(\zeta ,\eta ;t=0)$ and $D_0(\zeta ,\eta)\equiv \varphi _{2}(\zeta ,\eta ;t=0)$. We note that, due to the fator 
$\e^{-\zeta ^{2}-\eta ^{2}}$ in the coefficients $A(\zeta ,\eta) \sim D(\zeta ,\eta)$, the integrations on $\zeta$ and $\eta$ in Eqs. ($\ref{eq:PP}$) and ($\ref{eq:NP}$) are Gaussian type and will be always convergent.

Given an initial state, we can decompose it into a part of a positive and a negative parity, and hence, $\psi _{n}(\zeta ,\eta ;0)$ and $\varphi _{n}(\zeta ,\eta ;0)$  ($n=1, 2$) are known. Within Eqs. (\ref{eq:An})-(\ref{eq:Bn}) and (\ref{eq:Cn})-(\ref{eq:Dn}), all the expansion coefficient in the dynamical equations ($\ref{eq:PP}$) and ($\ref{eq:NP}$) can be systematically obtained.
Thus, the full dynamical properties of the physical observables are known.

\section{Conclusions}
In summary, we have studied the dynamical properties of the quantum Rabi model. 
The quantum Rabi model has the parity symmetry. As a result, the system keeps the same parities during the evolution of the states. 
Therefore the state vector at any time can be decomposed into a positive parity and a negative one, which, respectively, is expanded by
the superposition of the coherent states. The evolution equations for the positive and the negative parity are obtained, and 
all the expansion coefficient can be systematically obtained through the recurrence relation.

We would like to mention two advantages of the method. One is that we can systematically derive the time evolution of the dynamic properties without
using the Trotterization of the time interval. The other is that our method starts from the full quantum Hamiltonian of the Rabi model and does not rely on the adiabatic approximation or RWA on the Hamiltonian. The full dynamic properties of the wavepacket and thus the occupation number in the up and down atom level will be addressed in the future study.

\appendix

\section{Derivation of Eq. (\ref{FockCoh})}
\label{Sec:Fockcoh}

\begin{eqnarray}
\vert n\rangle &= \int\int e^{-\zeta^{2}-\eta^{2}}\vert\zeta+i\eta\rangle \langle\zeta+i\eta \vert n\rangle \frac{\mathrm{d}\zeta\mathrm{d}\eta}{\pi}\nonumber\\
&= \int\int e^{-\zeta^{2}-\eta^{2}}\vert\zeta+i\eta\rangle \left(\sum^\infty_{m=0}\frac{(\zeta-i\eta)^{m}}{\sqrt{m!}}\langle m\vert\right)\vert n\rangle \frac{\mathrm{d}\zeta\mathrm{d}\eta}{\pi}\nonumber\\
&= \int\int e^{-\zeta^{2}-\eta^{2}}\frac{(\zeta-i\eta)^{n}}{\sqrt{n!}}\vert\zeta+i\eta\rangle \frac{\mathrm{d}\zeta\mathrm{d}\eta}{\pi}\nonumber\\
&=\int\int e^{-\zeta^{2}-\eta^{2}}\sum_{l=0}^{n}\frac{\sqrt{n!}}{l!(n-l)!}(\zeta)^{n-l}(-i\eta)^{l}\vert\zeta+i\eta\rangle \frac{\mathrm{d}\zeta\mathrm{d}\eta}{\pi}.
\label{eq:aFock}
\end{eqnarray}
When $n=2m$,
\begin{eqnarray}
\vert 2m\rangle& =\frac{1}{2}\int\int e^{-\zeta^{2}-\eta^{2}}\sum_{l=0}^{2m}\frac{\sqrt{2m!}}{l!(2m-l)!}(\zeta)^{2m-l}(-i\eta)^{l}\nonumber\\
&~~~\times \left( \vert \zeta+i\eta\rangle +\vert -\zeta-i\eta\rangle  \right)\frac{\mathrm{d}\zeta\mathrm{d}\eta}{\pi},
\end{eqnarray}
and when $n=2m+1$,
\begin{eqnarray}
\vert 2m+1\rangle  &=\frac{1}{2}\int\int e^{-\zeta^{2}-\eta^{2}}\sum_{l=0}^{2m+1}\frac{\sqrt{(2m+1)!}}{l!(2m+1-l)!}
 (\zeta)^{2m+1-l}(-i\eta)^{l}\nonumber\\
&~~~\times \left( \vert \zeta+i\eta\rangle -\vert -\zeta-i\eta\rangle  \right)  \frac{\mathrm{d}\zeta\mathrm{d}\eta}{\pi}.
\end{eqnarray}

\section{Operation of the bosonic operators on the coherent state}
\label{Sec:oper}
In order to derive the recurrence formula for the time-dependent coefficients, we need to make use of the operation of the bosonic operators on the coherent state, 
\begin{eqnarray}
&~~~~a\int\int\psi(\zeta,\eta)[\vert\zeta+i\eta\rangle \pm\vert -\zeta-i\eta\rangle ]\mathrm{d}\zeta\mathrm{d}\eta\nonumber\\
&=\int\int\psi(\zeta,\eta)(\zeta+i\eta)[\vert\zeta+i\eta\rangle \mp\vert -\zeta-i\eta\rangle ]\mathrm{d}\zeta\mathrm{d}\eta,
\label{eq:acoh}
\end{eqnarray}
and,
\begin{eqnarray}
&~~~~a^\dagger \int\int\psi(\zeta,\eta)[\vert \zeta+i\eta\rangle +\vert -\zeta-i\eta\rangle ]\mathrm{d}\zeta\mathrm{d}\eta\nonumber\\
 &=a^\dagger \sum_{n}^{}\int\int\psi(\zeta,\eta)\vert n\rangle \langle n\vert [\vert\zeta+i\eta\rangle +\vert -\zeta-i\eta\rangle ]\mathrm{d}\zeta\mathrm{d}\eta
 \nonumber\\
 &=\sum_{n}^{}\int\int\psi(\zeta,\eta)\sqrt{n+1}\vert n+1\rangle \langle n\vert\left[\sum_{m}^{}\frac{(\zeta+i\eta)^{m}}{\sqrt{m!}}\vert m\rangle +\sum_{m}^{}\frac{(-\zeta-i\eta)^{m}}{\sqrt{m!}}\vert m\rangle \right ]\mathrm{d}\zeta\mathrm{d}\eta\nonumber\\
 &=\sum_{n}^{}\int\int\psi(\zeta,\eta)\sqrt{n+1}\vert n+1\rangle \left[\frac{(\zeta+i\eta)^{n}}{\sqrt{n!}}+\frac{(-\zeta-i\eta)^{n}}{\sqrt{n!}}\right ]\mathrm{d}\zeta\mathrm{d}\eta\nonumber\\
 &=\sum_{m}^{}\int\int\psi(\zeta,\eta)\sqrt{2m+1}\frac{2(\zeta+i\eta)^{2m}}{\sqrt{2m!}}\vert 2m+1\rangle \mathrm{d}\zeta\mathrm{d}\eta\nonumber\\
 &=\int\int\sum^\infty_{m=0}\sum_{l=0}^{2m+1}\frac{2m+1}{l!(2m+1-l)!} \int\int\psi(\zeta,\eta)(\zeta+i\eta)^{2m}\mathrm{d}\zeta\mathrm{d}\eta \nonumber\\
 &~~~~\times \e^{-\zeta_1^{2}-\eta_1^{2}}(\zeta_{1})^{2m+1-l}(-i\eta_{1})^{l} (\vert \zeta_{1}+i\eta_{1}\rangle -\vert -\zeta_{1}-i\eta_{1}\rangle )\frac{\mathrm{d}\zeta_1\mathrm{d}\eta_1}{\pi}.
 \label{eq:aaddcoh1}
\end{eqnarray}
Similarly,
\begin{eqnarray}
&~~~~a^\dagger \int\int\psi(\zeta,\eta)[\vert\zeta+i\eta\rangle -\vert -\zeta-i\eta\rangle ]\mathrm{d}\zeta\mathrm{d}\eta\nonumber\\
 &=\sum^\infty_{m=0}\sum_{l=0}^{2m+1}\frac{2m+2}{l!(2m+2-l)!} \int\int\psi(\zeta,\eta)(\zeta+i\eta)^{2m+1}\mathrm{d}\zeta\mathrm{d}\eta \nonumber\\
 &~~~\times \e^{-\zeta_1^{2}-\eta_1^{2}}(\zeta_{1})^{2m+2-l}(-i\eta_{1})^{l}
 (\vert\zeta_{1}+i\eta_{1}\rangle -\vert -\zeta_{1}-i\eta_{1}\rangle )\frac{\mathrm{d}\zeta_1\mathrm{d}\eta_1}{\pi}.
  \label{eq:aaddcoh2}
\end{eqnarray}
With Eqs. (\ref{eq:acoh}), (\ref{eq:aaddcoh1}) and (\ref{eq:aaddcoh2}), we obtain, 
\begin{eqnarray}
&~~~~a^\dagger a\int\int\psi(\zeta,\eta)[\vert\zeta+i\eta\rangle +\vert -\zeta-i\eta\rangle ]\mathrm{d}\zeta\mathrm{d}\eta\nonumber\\
 &=a^\dagger \int\int\psi(\zeta,\eta)(\vert\zeta+i\eta\rangle -\vert -\zeta-i\eta\rangle )\mathrm{d}\zeta\mathrm{d}\eta\nonumber\\
 &=\sum^\infty_{m=0}\sum_{l=0}^{2m+2}\frac{2m+2}{l!(2m+2-l)!}\int\int\psi(\zeta,\eta)(\zeta+i\eta)^{2m+2}\mathrm{d}\zeta\mathrm{d}\eta\nonumber\\
 &~~~\times \e^{-\zeta_1^{2}-\eta_1^{2}}(\zeta_{1})^{2m+2-l}(-i\eta_{1})^{l}
 (\vert\zeta_{1}+i\eta_{1}\rangle +\vert -\zeta_{1}-i\eta_{1}\rangle )\frac{\mathrm{d}\zeta_1\mathrm{d}\eta_1}{\pi},
\end{eqnarray}
and
\begin{eqnarray}
&~~~~a^\dagger a\int\int\psi(\zeta,\eta)[\vert\zeta+i\eta\rangle -\vert -\zeta-i\eta\rangle ]\mathrm{d}\zeta\mathrm{d}\eta\nonumber\\
 &=\sum^\infty_{m=0}\sum_{l=0}^{2m+1}\frac{2m+1}{l!(2m+1-l)!}\int\int\psi(\zeta,\eta)(\zeta+i\eta)^{2m+1}\mathrm{d}\zeta\mathrm{d}\eta\nonumber\\
 &~~~\times \e^{-\zeta_1^{2}-\eta_1^{2}}(\zeta_{1})^{2m+1-l}(-i\eta_{1})^{l}
 (\vert\zeta_{1}+i\eta_{1}\rangle -\vert -\zeta_{1}-i\eta_{1}\rangle )\frac{\mathrm{d}\zeta_1\mathrm{d}\eta_1}{\pi}.
\end{eqnarray}
The above equations will be used in the derivation of Eq. (\ref{eq:hofj}).


\ack
This work was supported by the NSF of China (Grant No. 11374266), the NSF of Zhejiang Province (Grant No. Z15A050001), 
and the Program for New Century Excellent Talents in University.


\section*{References}

\begin{thebibliography}{10}
\bibitem{books} Cohen-Tannoudji C, Dupont Roc J and Grynberg G 1992 {\it Atom-Photon Interactions} (New York, John Wiley \& Sons)
             \nonum Scully M O and Zubairy M S 2001 {\it Quantum Optics} (Cambridge, Cambridge University Press)

\bibitem{Rabi36} Rabi I I 1936 On the process of space quantization {\sl Phys. Rev.} \textbf{49} 324
             \nonum Rabi I I 1937 Space quantization in a gyrating magnetic field {\sl Phys. Rev.} \textbf{51} 652

\bibitem{cavity QED} Raimond J M, Brune M and Haroche S 2001 Manipulating quantum entanglement with atoms and photons in a cavity {\sl Rev. Mod. Phys.} {\bf 73} 565

\bibitem{haroche} Haroche S and Raimond J M 2006 {\it Exploring the Quantum: Atoms, Caviries and photons} (Oxford, Oxford University Press)

\bibitem{Dots} Leibfried D, Blatt R, Monroe C and Wineland D 2003 Quantum dynamics of single trapped ions {\sl Rev. Mod. Phys.} {\bf 75} 28

\bibitem{circuit QED} Roch N, Schwartz M E, Motzoi F, Macklin C, Vijay R, Eddins A W, Korotkov A N, Whaley K B, Sarovar M and Siddiqi I 2014 Observation of Measurement-Induced Entanglement and Quantum Trajectories of Remote Superconducting Qubits {\sl Phys. Rev. Lett.} {\bf 112} 170501
   \nonum Romero G, Ballester D, Wang Y M, Scarani V and Solano E 2012 Ultrafast Quantum Gates in Circuit QED {\sl Phys. Rev. Lett.} \textbf{108} 120501

\bibitem{wallraff1} Wallraff A, Schuster D I, Blais A, Frunzio L, Huang R S, Majer J, Kumar S, Girvin S M and Schoelkopf R J 2004 Circuit Quantum Electrodynamics: Coherent Coupling of a Single Photon to a Cooper Pair Box {\it Nature} {\bf 431} 162 

\bibitem{Bloch} Bloch I, Dalibard J and Zwerger W 2008 Many-body physics with ultracold gases {\sl Rev. Mod. Phys.} {\bf 80} 885

\bibitem{Schweber} Schweber S 1967 On the application of Bargmann Hilbert spaces to dynamical problems, {\sl Ann. Phys. (N.Y.)} {\bf 41} 205

\bibitem{Larson} Larson J 2007 Dynamics of the Jaynes–Cummings and Rabi models: old wine in new bottles {\sl Phys. Scr.} {\bf 76} 146

\bibitem{Nataf} Nataf P and Ciuti C 2010 Vacuum Degeneracy of a Circuit QED System in the Ultrastrong Coupling Regime {\sl Phys. Rev. Lett.} {\bf 104} 023601

\bibitem{Schiro} Schir\'{o} M, Bordyuh M, \"{O}ztop B and T\"{u}reci H E 2012 Phase Transition of Light in Cavity QED Lattices {\sl Phys. Rev. Lett.} {\bf 109} 053601

\bibitem{Wolf} Wolf F A, Vallone F, Romero G, Kollar M, Solano E and Braak D 2013 Dynamical correlation functions and the quantum Rabi model {\sl Phys. Rev. A} {\bf 87} 023835

\bibitem{Liberato} Liberato De S 2014 Light-Matter Decoupling in the Deep Strong Coupling Regime: The Breakdown of the Purcell Effect {\sl Phys. Rev. Lett.} {\bf 112} 016401

\bibitem{Henriet} Henriet L, Ristivojevic Z, Orth P P and Hur K Le 2014 Quantum dynamics of the driven and dissipative Rabi model {\sl Phys. Rev. A} {\bf 90} 023820

\bibitem{JC} Jaynes E T and Cummings F W 1963 Comparison of quantum and semi-classical radiation theories with application to beam maser {\it Proc. IEEE} {\bf 51} 89
      \nonum Shore B W and Knight P L 1993 The Jaynes-Cummings model {\it J. Mod. Opt} {\bf 40} 1195

\bibitem{Felicetti} Felicetti S, Rico E, Sabin C, Ockenfels T, Koch J, Leder M, Grossert C, Weitz M and Solano E Quantum Rabi model in the Brillouin zone with ultracold atoms arXiv:1606.05471v1

\bibitem{RWA} Bourassa J, Gambetta J M, Abdumalikov A A, Astafiev O, Nakamura Y and Blais A 2009 Ultrastrong coupling regime of cavity QED with phase-biased flux qubits {\sl Phys. Rev. A} {\bf 80} 032109

\bibitem{Ciuti} Ciuti C, Bastard G and Carusotto I 2005 Quantum vacuum properties of the intersubband cavity polariton field {\sl Phys. Rev. B} {\bf 72} 115303 

\bibitem{Gunter} G\"unter G, Anappara A A, Hees J, Sell A, Biasiol G, Sorba L, De Liberato S, Ciuti C, Tredicucci A, Leitenstorfer A and Huber R 2009 Sub-cycle switch-on of ultrastrong light-matter interaction {\sl Nature} \textbf{458} 178
\nonum Niemczyk T, Deppe F,	 Huebl H,	 Menzel E P, Hocke F, Schwarz M J, Garc{\'i}a-Ripoll J J, D. Zueco, H{\"u}mmer T, Solano E, Marx A and Gross R 2010 Circuit quantum electrodynamics in the ultrastrong-coupling regime {\sl Nature Physics} \textbf{6} 772
\nonum  Forn-D{\'i}az P, Lisenfeld J, Marcos D, Garc{\'i}a-Ripoll J J, Solano E, Harmans C J P M and Mooij J E 2010 Observation of the Bloch-Siegert shift in a qubit-oscillator system in the ultrastrong coupling regime {\sl Phys. Rev. Lett.} \textbf{105} 237001
\nonum Yoshihara F, Fuse T, Ashhab S, Kakuyanagi K, Saito S and Semba K 2016 
Superconducting qubit-oscillator circuit beyond the ultrastrong-coupling regime arXiv:1602.00415
\bibitem{Casanova} Casanova J, Romero G, Lizuain I, Garc{\'i}a-Ripoll J J and Solano E 2010 Deep strong coupling regime of the Jaynes-Cummings model {\sl Phys. Rev. Lett.} \textbf{105} 263603

\bibitem{Braak} Braak D 2011 Integrability of the Rabi model {\sl Phys. Rev. Lett.} \textbf{107} 100401

\bibitem{Chen} Chen Q H, Wang C, He S, Liu T and Wang K L 2012 Exact solvability of the quantum Rabi model using Bogoliubov operators {\sl Phys. Rev. A} \textbf{86} 023822

\bibitem{Zhong} Zhong H H, Xie Q T, Batchelor M T and Lee C H 2013 Analytical eigenstates for the quantum Rabi model {\sl J. Phys. A} \textbf{46} 415302

\bibitem{Maciejewski} Maciejewski A J, Przybylska M and Stachowiak T 2014 Analytical method of spectra calculations in the Bargmann representation {\sl Phys. Lett. A} \textbf{378} 3445

\bibitem{cont} Liu T, Feng M and Wang K L 2011 Vacuum-induced Berry phase beyond the rotating-wave approximation {\it Phys. Rev. A} {\bf 84} 062109 
\nonum Wang Y M, Du G and Liang J Q 2012 Geometric phases in qubit-oscillator system beyond conventional rotating-wave approximation {\it Chin. Phys. B} {\bf 21} 044207  

\bibitem{jonasBerry} Larson J 2012 Absence of Vacuum Induced Berry Phases without the Rotating Wave Approximation in Cavity QED {\it Phys. Rev. Lett.} {\bf 108} 033601

\bibitem{Hwang15} Hwang M J, Puebla R and Plenio M B 2015 Quantum Phase Transition and Universal Dynamics in the Rabi Model {\sl Phys. Rev. Lett.} {\bf 115} 180404

\bibitem{Hwang10} Hwang M J and Choi M S 2010 Variational study of a two-level system coupled to a harmonic oscillator in an ultrastrong-coupling regime {\sl Phys. Rev. A} {\bf 82} 025802

\bibitem{Fleit} Fleit M D, Fleck J A and Steiger A 1982 Solution of the Schr\"{o}dinger equation by a spectral method {\sl J. Comput. Phys.} {\bf 47} 412 

\bibitem{Pedernales} Pedernales J S, Lizuain I, Felicetti S, Romero G, Lamata L and Solano E 2015 Quantum Rabi model with trapped ions {\sl Sci. Rep.} \textbf{5} 15472  

\bibitem{Forndiaz} Forn-D\'{i}az P, Romero G, Harmans C J P M, Solano E and Mooij J E 2016 Broken selection rule in the quantum Rabi model arXiv:1510.03379

\end{thebibliography}
\end{document}